\documentstyle[emulateapj]{article}
\input psfig.sty
\submitted{Accepted by ApJ Letters}

\lefthead{Basu}
\righthead{Triaxiality of Cores}

\begin{document}

\newcommand{\beq}{\begin{equation}}
\newcommand{\eeq}{\end{equation}}
\newcommand{\refer}{\reference}
\newcommand{\ul}{\underline{\hspace{20pt}}}
\newcommand{\cs}{c_{\rm s}}
\newcommand{\cmc}{{\rm cm}^{-3}}
\newcommand{\Alf}{Alfv\'{e}n\ }
\newcommand{\sigv}{\sigma_v}
\newcommand{\Siglos}{\Sigma_{\rm los}}
\newcommand{\Blos}{B_{\rm los}}
\newcommand{\Bperp}{B_{\bot}}
\newcommand{\Nlos}{N_{\rm los}}
\newcommand{\mulos}{\mu_{\rm los}}
\newcommand{\Psiavg}{\langle\Psi\rangle}
\newcommand{\qavg}{\langle q \rangle}

\title{Magnetic Fields and the Triaxiality of Molecular Cloud Cores}

\author{Shantanu Basu}
\affil{Department of Physics and Astronomy, University of
Western Ontario, London, Ontario N6A 3K7, Canada; basu@astro.uwo.ca}

\begin{abstract}

We make the hypothesis that molecular cloud fragments are triaxial bodies
with a large scale magnetic field oriented along the short axis.
While consistent with theoretical expectations, this idea is supported
by magnetic field strength data, which show
strong evidence for flattening along the direction of the mean magnetic
field. It is also consistent with early submillimeter polarization data,
which show that the projected direction of the magnetic field is often
slightly misaligned with the projected minor axis of a molecular cloud core,
i.e., the offset angle $\Psi$ is nonzero.
We calculate distributions of $\Psi$ for various 
triaxial bodies, when viewed from a random set of viewing angles. 
The highest viewing probability always corresponds to $\Psi=0^{\circ}$, 
but there is a finite probability of viewing all nonzero $\Psi$, including even 
$\Psi =90^{\circ}$; the average offset typically falls in the range 
$10^{\circ}-30^{\circ}$ for triaxial bodies 
most likely to satisfy observational and theoretical constraints.

\end{abstract}

\keywords{ISM: clouds - ISM: magnetic fields - MHD - polarization}

\section{Introduction}

Observations establish that magnetic fields play an important role in
shaping the structure and dynamics of molecular clouds. Polarization
maps reveal that there is a magnetic field threading 
molecular clouds that is ordered over large scales (e.g., Vrba, Strom, \&
Strom 1976; Goodman et al. 1990; Schleuning 1998) 
and Zeeman measurements establish that measured field 
strengths are invariably strong enough to yield a mass-to-flux ratio
very close to the critical value (Crutcher 1999; hereafter C99), i.e.,
the magnetic energy approximately equals the gravitational binding energy. 
A problem with absorption maps of the polarization of background starlight
is that the densest regions (the cloud cores) are not sampled 
(Goodman et al. 1995).
However, new maps of 
polarized submillimeter emission (Matthews \& Wilson 2000; Ward-Thompson et al.
2000) do sample these regions and provide a crucial test of theoretical
scenarios involving magnetic fields; in particular, the relative alignment
of the projected magnetic field with the projected minor axis of the
core density distribution is an important diagnostic.

When making the above comparison, two important factors must be kept in
mind: 1) the measured relative alignment is only between two 
projected quantities in the plane of the sky, and care must be taken to
draw inferences about the relative orientation in the actual three dimensional
object, and  2) theoretical models of magnetically dominated clouds
(e.g., Mouschovias 1976; Lizano \& Shu 1989; Fiedler \& 
Mouschovias 1993),  
while predicting the important effect of flattening along the mean magnetic
field direction, are usually restricted (for numerical convenience) to the 
physically unnecessary assumption of axisymmetry. In fact, 
it has been known for a long time (Mestel 1965; Lin, Mestel,
\& Shu 1965) that gravity amplifies anisotropic structure, so that an
initially unstable fragment will first collapse along one dimension, forming
a sheet, which will subsequently break into elongated filamentary structures.
The presence of a large scale magnetic field can
enhance this effect, as the first stage of collapse is preferentially along
the mean field direction. The subsequent contraction lateral to the 
field will occur in qualitatively the same manner as described by Mestel (1965),
particularly if the cloud is magnetically supercritical, but even if it is 
subcritical, since the evolution is still driven by gravity (Mouschovias 1978; 
Langer 1978), but on the ambipolar diffusion (rather than free-fall) time scale.
Thus, the most general 
formation mechanism for a dense core in the magnetic field scenario yields 
an object flattened along the mean field direction and also having an
elongated structure in the lateral direction; such an object is triaxial.

The properties of the projected shapes of triaxial bodies have been 
investigated in the context of galaxy studies. Stark (1977) has proven the
important result that the isocontours of a triaxial body seen in projection are
elliptical. Binney (1985) has extended this analysis, and his results
can be applied to find the distribution of angular offsets
$\Psi$ between the projected field direction and the projected minor
axis of a triaxial body, given that the body is primarily flattened along
the direction of the mean magnetic field.

In \S\ 2, we review observational evidence to establish that molecular 
cloud fragments are ubiquitously flattened along the mean magnetic 
field direction, and in \S\ 3 calculate the distribution of $\Psi$ for various
triaxial bodies. In \S\ 4 we discuss how the results relate
to current observations and to other theoretical scenarios presented 
in the literature.

\section{Cloud Flattening along Magnetic Field Lines}

The magnetic field strength data of C99 also 
reveals that the correlation of the line-of-sight field 
strength $\Blos$ with the density $\rho$ is in apparent 
agreement with models of preferential flattening
along the magnetic field ($B \propto \rho^{1/2}$; Mouschovias 1976) rather than isotropic contraction ($B \propto \rho^{2/3}$).
Figure 1a (equivalent to Fig. 1 of C99) plots (in logarithmic form) the 
confirmed magnetic field strength detections of C99 against the 
measured number densities $n$. 
The solid line is the least squares best fit, with
slope $0.47 \pm 0.08$, and chi-squared error statistic $\chi^2 = 2.36$.
The correlation coefficient is 0.84. 

We seek an even better correlation for the C99 data as follows.
Consider a cloud that is flattened along the mean magnetic field direction.
In this direction, force balance between an isotropic pressure and self-gravity
yields
\beq
\rho_0 \, \sigv^2 = \frac{\pi}{2} G \, \Sigma^2,
\label{forcebal}
\eeq
where $\rho_0$ is the density at the midplane, $\sigv$ is the total
(thermal and non-thermal) one-dimensional velocity dispersion, $\Sigma$
is the column density of the cloud, and $G$ is the gravitational constant.
Here we assume that the effect of an external confining pressure is negligible.
Furthermore, the relation
\beq
\frac{\Sigma}{B} \equiv \mu \, (2 \pi G^{1/2})^{-1}
\label{defmu}
\eeq
defines the dimensionless ratio $\mu$ of the mass-to-flux ratio to the critical
value $(2 \pi G^{1/2})^{-1}$ for a disk (Nakano \& Nakamura 1978).
Combining equations (\ref{forcebal}) and (\ref{defmu}) yields
\beq
B = (8 \pi c_1)^{1/2} \; \frac{\sigv \rho^{1/2}}{\mu}
\label{brho}
\eeq
(see Mouschovias 1991 for a similar relation), where $c_1 \, (\gtrsim 1)$ is an 
undetermined proportionality constant between $\rho_0$ and the mean 
density $\rho$. Clearly, the correlation in Figure 1a is imperfect because 
variations in $\sigv$ and $\mu$ are unaccounted for, 
and the observed quantity $\Blos= B \cos \theta$ (where $\theta$ is the 
angle between the field direction and the line of sight) does not
correspond exactly to $B$. We improve the correlation by accounting for 
variations in the observable quantity $\sigv$. Figure 1b plots $\Blos$ 
versus $\sigv \, n^{1/2}$ in
logarithmic form; equation (\ref{brho}) predicts a slope of 1 to the extent 
that $\mu$ is constant from cloud to cloud. We use values of 
$\Delta v = \sigv (8 \ln 2)^{1/2}$
listed in Table 1 of C99. The correlation coefficient in this plot is 0.95.
The solid line is the best least squares fit and has slope $1.00 \pm 0.09$, and
$\chi^2 = 0.79$. The dashed line is the theoretical relation (\ref{brho}) for
$\sqrt{c_1}/\mu = 1$. The two lines coincide if $\sqrt{c_1}/\mu=0.8$. 
This exceptional correlation                                  
leads us to the conclusion that clouds (and cloud fragments) are
flattened (at least moderately) along the magnetic field direction,
even in the presence of
turbulent motions, due to the dynamically important roles of self-gravity and
the mean magnetic field.
We note that the correlation found in Figure 1b is equivalent to that 
presented by Myers \& Goodman (1988) in their Figure 1 (see also Mouschovias \& 
Psaltis 1995) based on earlier 
magnetic field data. We also note that our fit to the magnetic field data 
is quite general; it does not require axisymmetry, or
that the \Alf speed $v_A = B/(4 \pi \rho)^{1/2}$ be constant from one cloud to
another. However, equation (\ref{brho}) implies that
$\sigv/v_A = (2 c_1)^{-1/2} \mu$, so that the \Alf mach number
may be approximately constant, if non-thermal motions dominate thermal
motions and $\mu$ does not vary much from one cloud to another.

\psfig{file=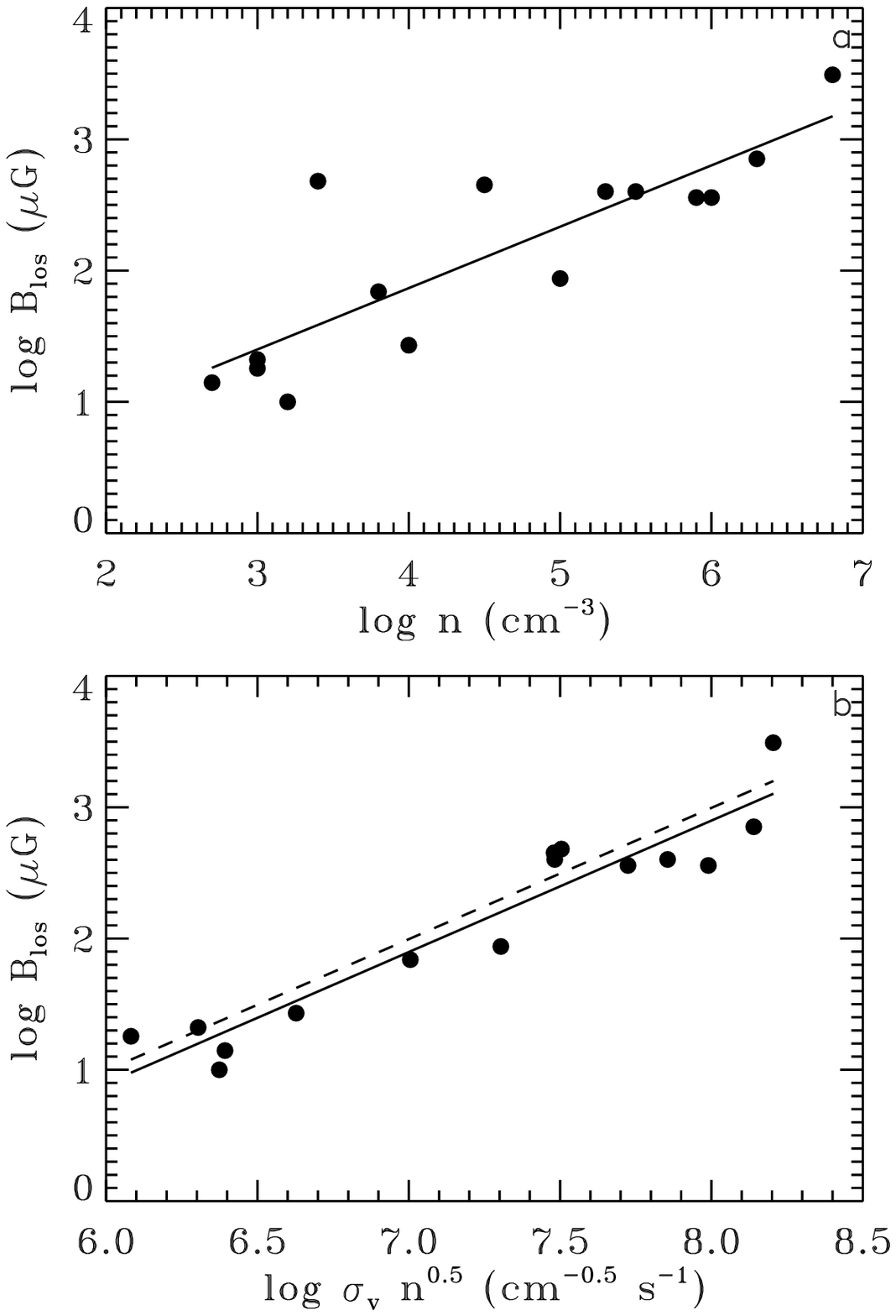,width=8.5cm}
\smallskip
\smallskip
\smallskip\noindent {\sl Fig.~1~}\
a) Plot of $\log \Blos$ versus $\log n$ from the C99 data set.
The solid line is the best fit
(see text). The correlation coefficient is 0.84.
b) Plot of $\log \Blos$ versus $\log \sigv n^{1/2}$ for the
same data. The correlation coefficient is 0.95. The solid line is the best
fit, and the dashed line is a theoretical relation (eq. [\ref{brho}] for
$\sqrt{c_1}/\mu = 1$).

\section{Projection of Triaxial Bodies}

Let $z$ be the direction of the mean magnetic field. Since the mean field
contributes cloud support against gravity only in the lateral ($x$ and $y$) 
directions, but the cloud need not be axisymmetric in the $x-y$ plane, the
most general cloud configuration is a triaxial body with
the shortest dimension along the $z-$axis.
The simplest mathematical description of such a body is 
\beq
x^2 + \frac{y^2}{\zeta^2} + \frac{z^2}{\xi^2} = a^2,
\eeq
where $a$ is a constant, and $1 \geq \zeta \geq \xi$.
Here we follow the geometric analysis of Binney (1985) to calculate 
projected quantities when the object is viewed from
an observing angle $(\theta, \phi)$ (where the 
angles are defined on an imaginary viewing sphere and have their usual
meaning in a spherical coordinate system). Defining the quantities
\beq
A \equiv \frac{\cos^2\theta}{\xi^2} \left(\sin^2 \phi + \frac{\cos^2 \phi}{\zeta^2} \right) + \frac{\sin^2\theta}{\zeta^2},
\eeq
\beq
B \equiv \cos\theta \sin 2\phi \left ( 1 - \frac{1}{\zeta^2} \right) \frac{1}{\xi^2},
\eeq
and
\beq
C \equiv \left( \frac{\sin^2\phi}{\zeta^2} + \cos^2\phi \right) \frac{1}{\xi^2},
\eeq
it can be shown that the projection in the sky of the triaxial body is an 
elliptical body of apparent axial ratio
\beq
q = \left( \frac{A + C - D}{A + C + D} \right)^{1/2},
\label{qratio}
\eeq
where $D \equiv \sqrt{(A-C)^2 + B^2}$.
Furthermore, the quantity
\beq
\psi = \frac{1}{2} \arctan \left( \frac{B}{A-C} \right)
\label{smallpsi}
\eeq
can be used to find the absolute value of the angular offset between the 
projected $z$ axis 
and the apparent minor axis of the projected ellipse. 
Letting $p = (A-C) \cos 2\psi+ B \sin 2\psi$, the required positive definite 
angle is
\beq
\Psi = \left\{ \begin{array}{ll}
               |\psi| &  \mbox{for $p \leq 0$} \\
               \pi/2 - |\psi|  & \mbox{for $p > 0$}. 
              \end{array}
       \right.
\label{bigpsi}
\eeq

Since $\Psi$ corresponds to the offset between the projected
magnetic field $\Bperp$
and the projected minor axis of a core, we calculate the
distribution of observed offsets $\Psi$ using a Monte Carlo simulation.
A large number $N = 10^6$ of randomly oriented viewing angles $(\theta,\phi)$
are used to generate a distribution of $\Psi$ for a given set of
intrinsic axial ratios $\zeta$ and $\xi$ of a triaxial body.
Figure 2 shows, for each of three cases $\xi =0.1,0.3,$ and 0.5, the 
probability distribution function $f$ versus $\Psi$ for three different 
values of $\zeta$, as well as the dependence of
the mean value $\Psiavg$ of the distribution and the mean axial ratio
$\qavg$ on $\zeta$. The lower limit
of $\zeta=\xi$ corresponds to a prolate object and the upper limit $\zeta
=0.9$ is very nearly an oblate object. 

Our calculated distributions show the robust result that there is always a
peak in $f$ at $\Psi=0$, so that it is the most probable value, but 
that there is a long tail towards all nonzero angles, including 
$\Psi = 90^{\circ}$. Figures 2b, 2d, and 2f show how 
$\Psiavg$ varies for clouds of different triaxial ratios;
it becomes smaller as a cloud goes toward the oblate limit. For a
perfectly oblate cloud $(\zeta=1)$, there would be no distribution, 
as all viewing angles yield $\Psi=0$.
The distributions for prolate clouds ($\zeta=\xi$)\footnote{Such 
clouds are distinct from existing models of prolate clouds 
(Tomisaka 1991; Fiege \& Pudritz 2000) in which the
mean magnetic field lies along the major axis; here the mean field lies
along the minor axis. The cloud can be prolate ($\zeta=\xi$) by chance, but 
there is no intrinsic symmetry that favors such a configuration, and
the pressure exerted by the mean field makes it likely that $\zeta > \xi$
for most clouds.}
have the largest possible values for $\Psiavg$. The values of $\qavg$ 
illustrate which triaxial models may be most likely. The crosses
highlight the values of $\zeta$ for which $\qavg$ is in the range $0.5 - 0.65$ 
that is consistent with observations of dense cores (e.g., Myers et al. 1991;
Jijina, Myers, \& Adams 1999). 

While data sets that measure $q$ and $\Psi$ for a large number of cores 
will ultimately settle the
values (and distributions) of $\zeta$ and $\xi$, we do not currently favor 
highly flattened cases $\xi \approx 0.1$ due to 1) the observational 
result that highly elongated cores are rarely observed, and 2) the 
theoretical result that flattening along the magnetic field yields cores 
of axial ratio
$\approx 0.25 - 0.33$ (see Ciolek \& Basu 2000 and references therein). 
Hence, the $\xi= 0.3$ curves represent likely scenarios, although
the $\xi= 0.5$ curves cannot be ruled out observationally. The observational
constraint on $\qavg$ also allows the widest plausible range of $\zeta$ 
values when $\xi = 0.3$, rather than imply near-oblate (for
$\xi=0.1$) or near-prolate (for $\xi=0.5$) configurations.
For $\xi$ in the range $0.3-0.5$, clouds with
$\qavg$ in the range $0.5-0.65$ have $\Psiavg$ in the range
$\simeq 10^{\circ} - 30^{\circ}$.

To illustrate the point that no single observation can establish the 
relative orientation of the field and the minor axis of the intrinsic
three-dimensional object, we present in Figure 3 a view of a triaxial
body with $(\xi,\zeta) = (0.3,0.6)$ from three positions on the
viewing sphere. This illustrates that all possible orientations of $\Bperp$
with the projected minor axis are possible, although the first panel 
shows the single most likely orientation and the middle panel is
closest to the average orientation $\Psiavg = 20^{\circ}$ for this case.

\section{Discussion}

Our predicted probability distributions, with their correlation toward $\Psi=0$,
are in good agreement with 
early submillimeter polarimetry of dense regions.
Matthews \& Wilson (2000) measure three distinct
regions in OMC-3, of which two imply near perfect alignment of $\Bperp$
with the projected minor axis, and one has $\Psi \gtrsim 
30^{\circ}$. Ward-Thompson et al. (2000) infer $\Bperp$ in three dense cores, 
L1544, L183, and L43, and also find it to be correlated with the
projected minor axes. They measure $\Psi = 29^{\circ}, 34^{\circ}$, and 
$44^{\circ}$ for the three cores, respectively, although the last measurement
is complicated by a nearby outflow and a weak polarization signal 
toward the core center.
For L1544, the nonzero $\Psi$
implies that the axisymmetric assumption in the magnetic model for that core
developed by Ciolek \& Basu (2000) needs to be relaxed, at least moderately. 
However, since that model does already account for 
flattening along the field lines, the magnetic field strength predictions
would likely not change substantially.

Future large data sets of the offset angle $\Psi$, as submillimeter polarization
measurements become common, will properly test the distributions of 
$\Psi$ presented in this paper. Such data will provide a 
clear observational test between the scenario presented here
(which is already supported by the Zeeman data), in which
gravity and the mean magnetic field play an important role in shaping
core structure (even in the presence of turbulence) and those scenarios
in which turbulent compression is dominant, leading to the claim 
(Ballesteros-Paredes, Vazquez-Semadeni, \& Scalo 1999) that there is no
correlation between the magnetic field direction and the cloud elongation.

\begin{acknowledgements}

I thank the anonymous referee for comments which helped improve the paper.
This research was supported by a grant from the Natural Sciences and
Engineering Research Council of Canada.

\end{acknowledgements}

%\clearpage

%\section*{Captions to Figures}

\psfig{file=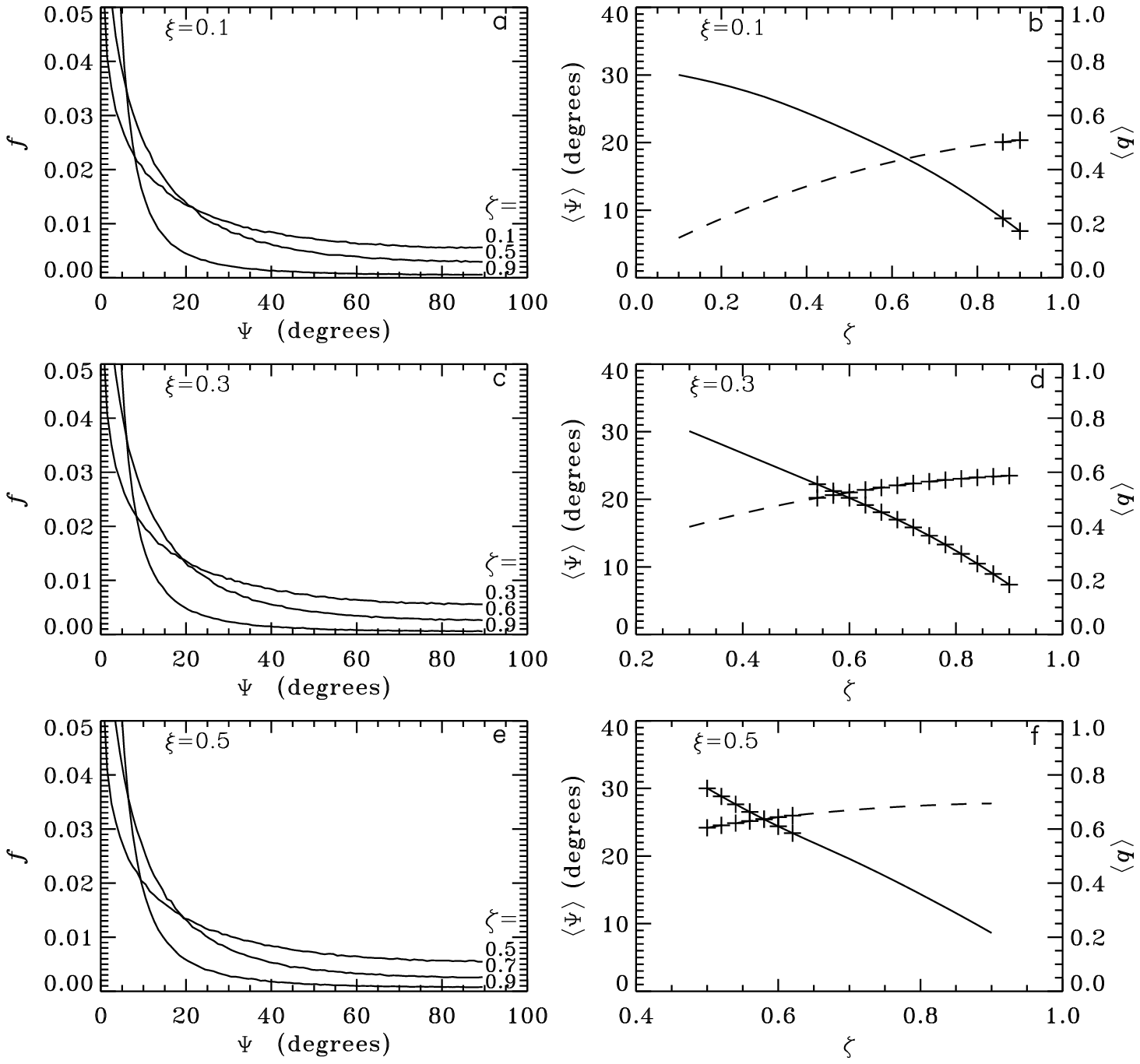,width=5.5in}
\smallskip
\vspace{0.1in}
\smallskip\noindent {\sl Fig.~2~}\
Probability distribution function $f$ versus angle $\Psi$ for
various values of $(\xi,\zeta)$, as
well as the dependence of mean values $\Psiavg$ and $\qavg$ on $\zeta$ 
for individual values of $\xi$. a) $f$ versus $\Psi$ for $\xi=0.1$ and
$\zeta = 0.1,0.5,0.9$. b) $\Psiavg$ (solid line) and $\qavg$ (dashed line)
versus $\zeta$ for $\xi=0.1$. c) Same as a) but $\xi=0.3$ and $\zeta = 
0.3, 0.6, 0.9$. d) Same as b) but $\xi=0.3$. e) Same as a) but $\xi=0.5$
and $\zeta = 0.5, 0.7, 0.9$. f) Same as b) but $\xi=0.5$.
In b), d) and f), the crosses highlight the region in which $\qavg = 0.5-0.65$,
consistent with observations.

\psfig{file=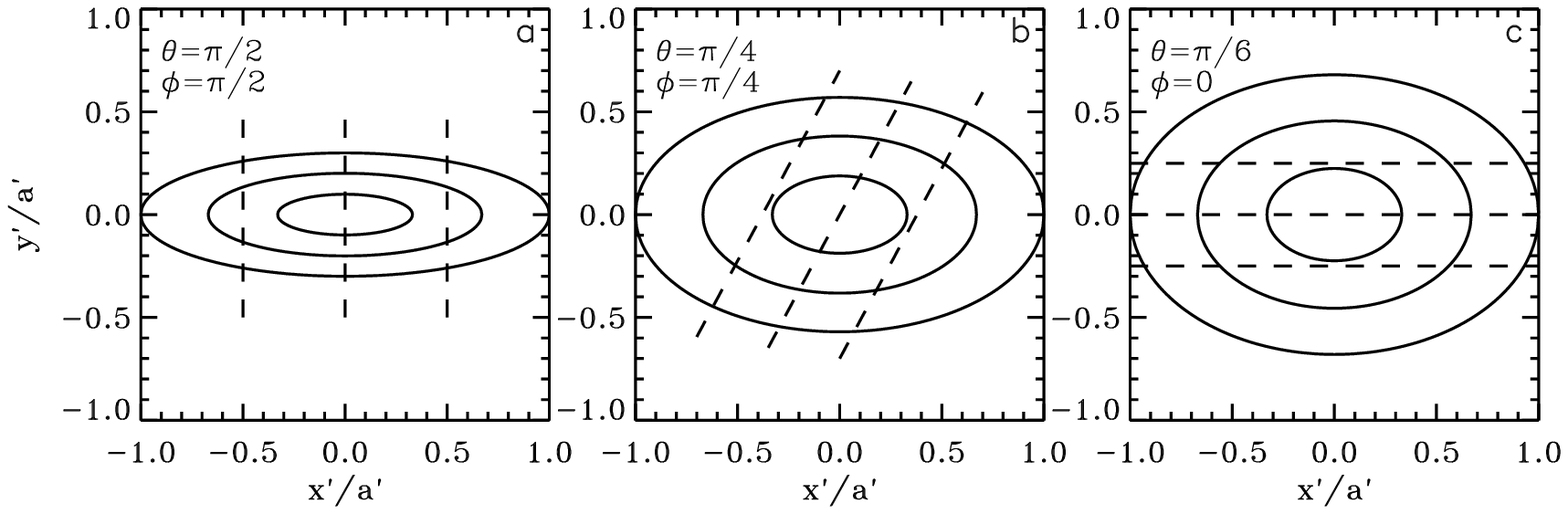,width=6.5in}
\smallskip
\smallskip\noindent {\sl Fig.~3~}\
Simulated contours (solid lines) and mean polarization direction 
(dashed lines) for a triaxial
body with axial ratios $(\xi,\zeta) = (0.3,0.6)$ seen from three
sets of viewing angles $(\theta, \phi)$. The
The $x'$ and $y'$ axes are in the plane of the sky and are chosen to lie 
along the projected major and minor axes, and $a'$ is the apparent 
semi-major axis. The dashed lines 
lie along the projected direction of the intrinsic $z$ axis. 
a) $(\theta, \phi) = (\pi/2,0)$, yielding
an apparent axis ratio $q = 0.3$ and offset angle $\Psi = 0^{\circ}$.
b) $(\theta, \phi) = (\pi/4,\pi/4)$, yielding $q = 0.57$ and 
$\Psi = 28^{\circ}$.
c) $(\theta, \phi) = (\pi/6,0)$, yielding $q = 0.68$ and  
$\Psi = 90^{\circ}$.

\end{document}